# Title

Declines in research funding and science ecosystem fragility

# Authors

Anbang Du[1]*, Beining Zhang[1], Rifat Atun[2,3,4]#, Michael G Head[5,6,7]#, Markus Brede[1]#

1. School of Electronics and Computer Science, University of Southampton, Southampton, UK
2. Health Systems Innovation Lab, Harvard T H Chan School of Public Health, Harvard University, Boston, MA, USA
3. Faculty of Medicine, Imperial College London, London, UK
4. Harvard Medical School, Harvard University, Boston, MA, USA
5. School of Primary Care, Population Sciences and Medical Education, Faculty of Medicine, University of Southampton, Southampton, UK
6. Fred N Binka School of Public Health, University of Health and Allied Sciences, Volta Region, Ghana
7. School of Medicine, University for Development Studies, Tamale, Ghana

*corresponding author, ad1u21@soton.ac.uk
# *joint senior author*

# Summary

Scientific knowledge advances through within-country and cross-border scientific activities and collaborations, influenced by funding and strength of research enterprise.[1] Sudden declines in research funding, for example from Federal sources in the United States (US) 2024-25,[2–4] adversely impact on scientific collaboration. How rapid declines in funding affect the science enterprise and the magnitude of impact need to be analysed.

Past studies have modelled the global scientific system as complex collaborative networks of entities and studied its topology and dynamics.[5–9] However, these studies have not undertaken compensation analysis to real-world shocks that have produced rapid declines in scientific research funding.

In this study we examine the effect of the sharp declines in the US Federal funding on cancer science research enterprise globally. We model the cancer science ecosystem as a 5-layer multiplex network of collaborative linkages between 233 countries and territories in grants and clinical trial co-investigations, paper co-authorships, co-inventions and patent co-ownerships. We quantify information flow in the multiplex system through network efficiency.[10,11]

Proposing a framework for compensation analysis, we show that sharp declines in US Federal funding for research degrade global information exchange in science, imposing outsized compensatory burdens on country groups such as the European Union (EU) and BRICS (Brazil, Russia, India, China and South Africa). However, we also show that if other countries provide more support for international collaborations, there is an opportunity to remodel the cancer science system to be more resilient to future shocks.

## Introduction

The structure of global science[12] is shaped by the geo-political and economic distribution of activity, resources and status. Dominant actors project strong influence on who participates in, and benefits from, research, and from where value from research is created and captured. Using statistical mappings and social network analyses [13–15] the resulting hierarchies and inequalities[1] have been examined through individual stages of the scientific process, from research funding[5,16], to publications[6], clinical trials[17,18] and patents[19].

Global investments in research and development (R&D) and research capabilities are typically concentrated in a small set of high-income economies[5]. For health research, investments are often not well correlated with disease[16]. Further, researchers in low- and middle-income countries (LMICs) are underrepresented in lead authorship positions in publications[6] or leadership of clinical trials[17,18]. Innovation capacity and economic benefits generated from intellectual property are overwhelmingly concentrated in a few high-income economies.[19] However, few studies have examined structural inequalities that exist in each stage of the scientific process.

In this paper we examine international cancer science ecosystem that receives large amounts of public funding. In 2023, there were 10.6 million deaths worldwide from cancer, second only to cardiovascular diseases.[20] Morbidity and mortality levels for cancer are rising rapidly in LMICs, where there is suboptimal access to healthcare services for cancer.[21]

We examine cancer science research and outputs as a value chain, where resources are acquired through grants, research findings are generated through studies such as clinical trials, knowledge is produced through publications, and value is created and captured through usage of knowledge, invention and intellectual property (IP) in the form of patents.

We model the international cancer science ecosystem as a 5-layer multiplex network of collaborations between 233 countries and territories, in grants, trial co-investigations, publication co-authorships, co-inventions and patent co-ownerships, to study the extent to which country actors are included or absent at each stage of the value chain.

Past studies of real-world networked systems have evaluated functional performance through quantifying network efficiency.[10,11,22] Here, we adopt this framework using network efficiency to monitor the information exchange across 233 countries and territories to study structural inequalities along the cancer science system value chain.

Using this model of the cancer science ecosystem we then examine the fragility of the ecosystem, considering fluctuations in activity from each country related to sudden changes in research funding. In particular, we analyze the effect of the sharp declines in R&D funding for cancer science from federal research funding agencies in the United States (US), which has the largest R&D hubs globally, hosting several of the world's major research funders, and forms the core of publication, IP, and trial capacity.[5,16] Changes in the US political administration in 2024-2025 was followed by an estimated 40.6% reduction in federal funding from research agencies such as the National Institutes of Health (NIH) in the fiscal year 2026.[2–4]

We also examine the possible compensation strategies that could be pursued by other actors, including by country groups such as the European Union (EU) and BRICS (Brazil, Russia, India, China and South Africa). We explore if the declines in funding from the US provide an opportunity to build a more resilient science ecosystem that is less dependent on a small number of actors.

# Results

## Inequality in global cancer science collaboration ecosystem

We start by encoding the collaboration in the value chain of the international cancer science ecosystem as a 5-layer multiplex (ML), where nodes represent countries and links fractional counts of collaboration activities (see method) in the layers of grants, publications, clinical trials, inventions, and patents (Fig.1(a)). To obtain a summary representation of overall collaboration intensity, we normalise individual layers of the multiplex to have identical average link weight and then aggregate links across layers[23], see Fig.1(b). In each layer and in the aggregate, the community of countries that can share knowledge, experience, skills, or value can be represented by the set of countries that can reach each other through connected paths, i.e. the largest connected component (LCC).

We argue that the more important role countries play in the cancer collaboration ecosystem the more layers countries exchange information in. Colouring nodes by the number of layers for which they belong to LCCs allows identification of central countries in the cancer science system (see Fig.1(b)). We note that the US and China (CN) belong to LCCs in all five layers and are thus in the core of the cancer science multiplex, while countries with small economies, such as Congo (CG), Tonga (TO) and Malta (MT), only partake in the LCC of the paper-related layer and are thus peripheral. We also note that these core countries tend to be well-connected topologically and contribute a disproportionately large share of link weight in each individual layer. (Supplementary Figure 1, 2, 3 and 4)

Having observed the structure of collaborations in the international cancer science ecosystem, we next explore how well countries of different income level are involved in each stage of the value chain. For this purpose, for each income level we calculate the fraction of countries that are part of LCCs belonging to grants, publications, clinical trials, invention and patenting (Fig1(c)). For all country income groups, we observe low involvement rates in the grant layer, a peak at the publication layer, and a drop-off towards the clinical trial, invention, and patent layers. In Fig1(c), we see that low-income countries (LICs) are often only part of the LCC of the publication layer, indicating little to no participation in grant, clinical trial, invention, and patent layers. For example, many LICs, such as South Sudan (SS), Burundi (BI), Eritrea (ER) and Somalia (SO), are only present in the LCC of the publication layer but not of the grant or patent layers. (see Supplementary Table 7)

As LICs often only contribute to the publication layer, further insights about their more detailed standing can be gained by analyzing authorship positions[8]. For this purpose, we included a comparison of LCC sizes between paper co-authorship networks based on all authors or restricted to first and last authorship positions. In the latter case, we find substantially reduced involvement of LICs and thus low representation in prestigious authorship positions. (Supplementary Figure 5)

Inequality is also represented in the different positioning of countries relative to information exchange on the multiplex. To further quantify this, we consider information exchange between all pairs of countries and count proportions of shortest paths routed through countries. When constructing shortest paths, a country can be directly connected to its partners. Alternatively, to reach another country, paths might be required to be routed through further intermediary countries. While communications in the former case are direct, they might require facilitation of information exchange from other countries in the latter case. Fig1(d) gives the fraction of direct shortest paths (blue bars), as well as the fractional involvement in intermediary paths (orange bars) for countries subdivided by income groups.

We note that the proportion of direct shortest paths is highest for HICs and steeply drops off towards LICs. This means that knowledge exchange involving HICs is easier and not reliant on other countries. For instance, US related links are mostly direct shortest paths (Supplementary Figure 6) whereas LICs typically have no direct connections. Additionally, HICs involvement in intermediary paths is very strong while LICs are severely under-represented in intermediary roles. HICs thus serve as major facilitators for global knowledge flows, with the US forming a dominant bridge (Supplementary Figure 6). By contrast, LICs are again found at the periphery of the knowledge exchange.

Country positioning also depends on the relative location of strong and weak links. To further probe the arrangement of connections by weight, we next decompose the aggregated network into cores of countries connected by strong links. This is done by starting with the original aggregated network and then gradually removing connections weaker than a given threshold. Subsequently, we analyse the economic characteristics of countries belonging to the LCCs of the remaining network. We find that the average GDP, GDP per capita, and R&D percentage of GDP of countries strongly increase with the threshold (Fig1(e)). The finding demonstrates the existence of a tightly knit core of rich countries connected by strong links. These countries also tend to participate in each layer of the value chain (Supplementary Table 7).

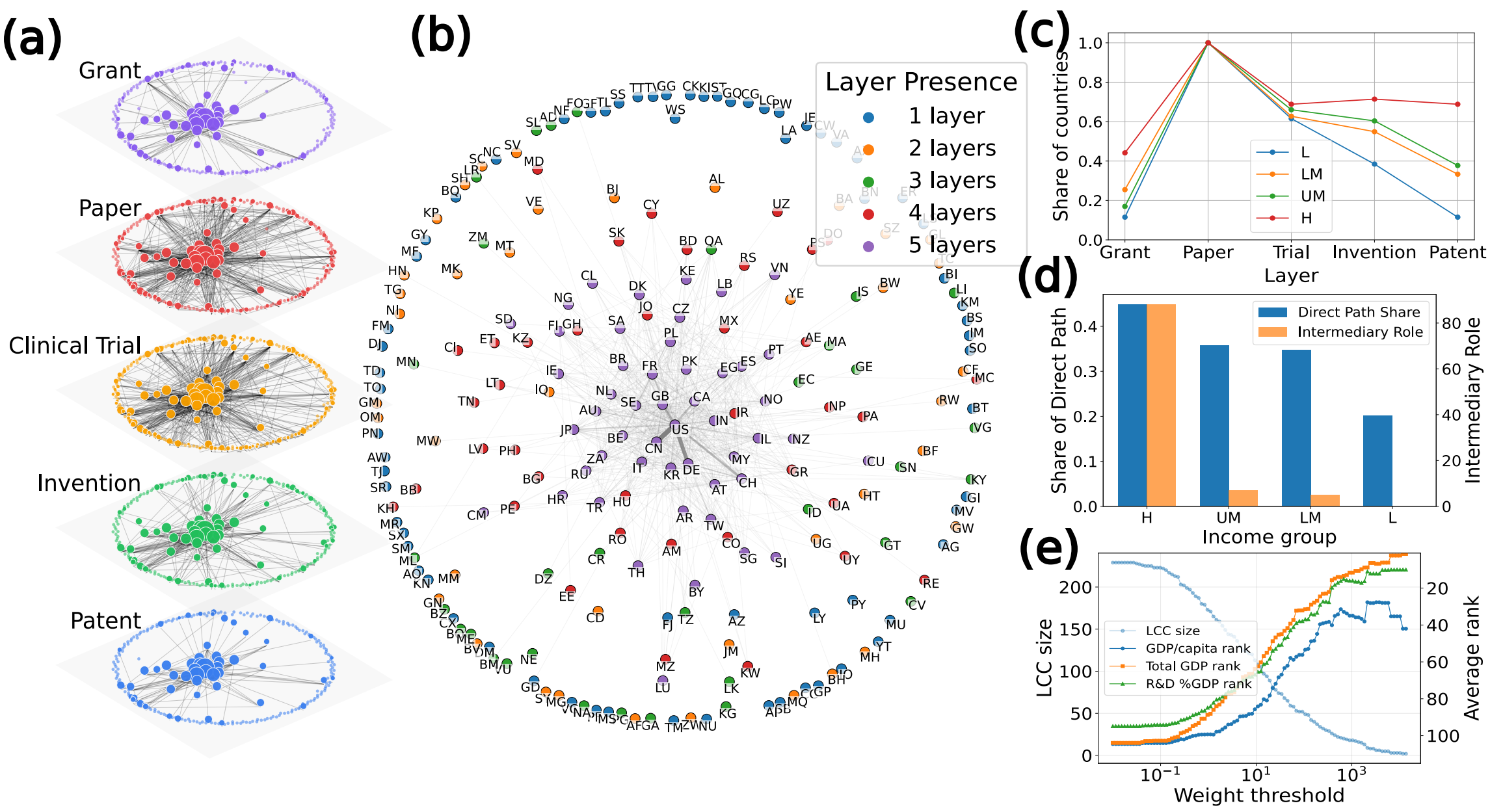

Fig1. Global cancer science collaboration ecosystem and analysis of multiplex network representation of grant co-investigations, paper co-authorships, clinical trial co-investigations, co-inventions, and co-patents. (a) Visualisation of the 5 layers of the multiplex network. Nodes are 233 countries and regions, and edges represent fractional counts[24] of collaborations. (b) Visualisation of the aggregated network of the five-layer multiplex network. For each layer, LCCs were constructed. Node color indicates the number of LCCs the node is part of. (c) Fraction of countries of different income levels (low (L, n=26), lower-middle (LM, n=51), upper-middle (UM, n=53), and high (H, n=78)) that contribute to the LCCs of each layer. (d) Share of direct (blue bars) and intermediary (orange bars) paths for countries of different income groups. (e) Dependence of the size of the LCC of the aggregated network and the average income rankings of GDP, GDP per capita, and R&D percentage of GDP from 2020 of countries in the LCC on weight thresholds under decomposition where links with weight lower than the threshold are removed.

## Fragility of the aggregated network

Having observed the inequality in the international cancer science ecosystem multiplex, we describe the fragility of the system, subject to activity reduction from key actors. in this context, reductions correspond to decreases in collaborative publishing and grant funding dedicated to international collaborations, as well as less international co-invention, or international clinical trials.

We model these activity reductions by scaling link weights on the multiplex. Similar to previous work in other contexts[10,11,22] we then measure the dependence of the global efficiency of the aggregated network on the percentage of link weight removed. We also probe fragility in response to countries' activities by scaling down the link strengths of each country/group (Fig2(a)). As expected, we observe that the overall communication efficiency drops as more activity is removed. Interestingly, efficiency responses differ between actors. For instance, compared to other major groupings, such as the G7, BRICS, or the EU, a reduction of the US activity causes a substantially stronger decay in the global efficiency. In contrast, communication is relatively resilient to activity reduction of non-US G7 countries, the EU, and BRICS. The result highlights the strong dependence of information flows on US activity and points to relative redundancy to activity reduction in other big HICs.

The dependence of the cancer science ecosystem on US is further emphasised by results in Fig.2(b), where we show efficiency losses in information exchange, subject to the removal of entire countries from the cancer science ecosystem. Whereas the removal of some leading actors such as China or Germany has a less than 10% impact, the removal of the US causes a dramatic loss of more than 50% of efficiency. One might argue that these findings are explained by the size of the contribution of the US to the science ecosystem. However, in Fig.2(b) we also probe the differential efficiency loss subject to removing one unit of activity from countries and again note that the ecosystem is highly dependent on the US.

Losses of efficiency are related to countries' positioning in knowledge exchange, which may be possible through direct connections between countries or require routing through third party countries to enable a path. We quantify this by measuring countries involvement in direct and intermediary paths. Results shown in Fig2(c) highlight that led by the US, big HICs tend to be connected via direct paths and have a dominating involvement as intermediaries, thus being enablers of communication between other countries. For example, we note that 108 out of 228 connections between the US and other countries are direct, whereas countries like Brazil (BR) and Fiji (FJ) have only two direct connections. Similarly, the US serves as intermediary for 54% of worldwide paths, whereas, Tonga and South Sudan are not part of any intermediary connections. This positioning also has implications for network resilience. For example, the dramatic losses resulting from a US withdrawal are explained by the US's share of 45% of intermediary paths. The lesser impact of other countries relates to their much smaller involvement in indirect communication.

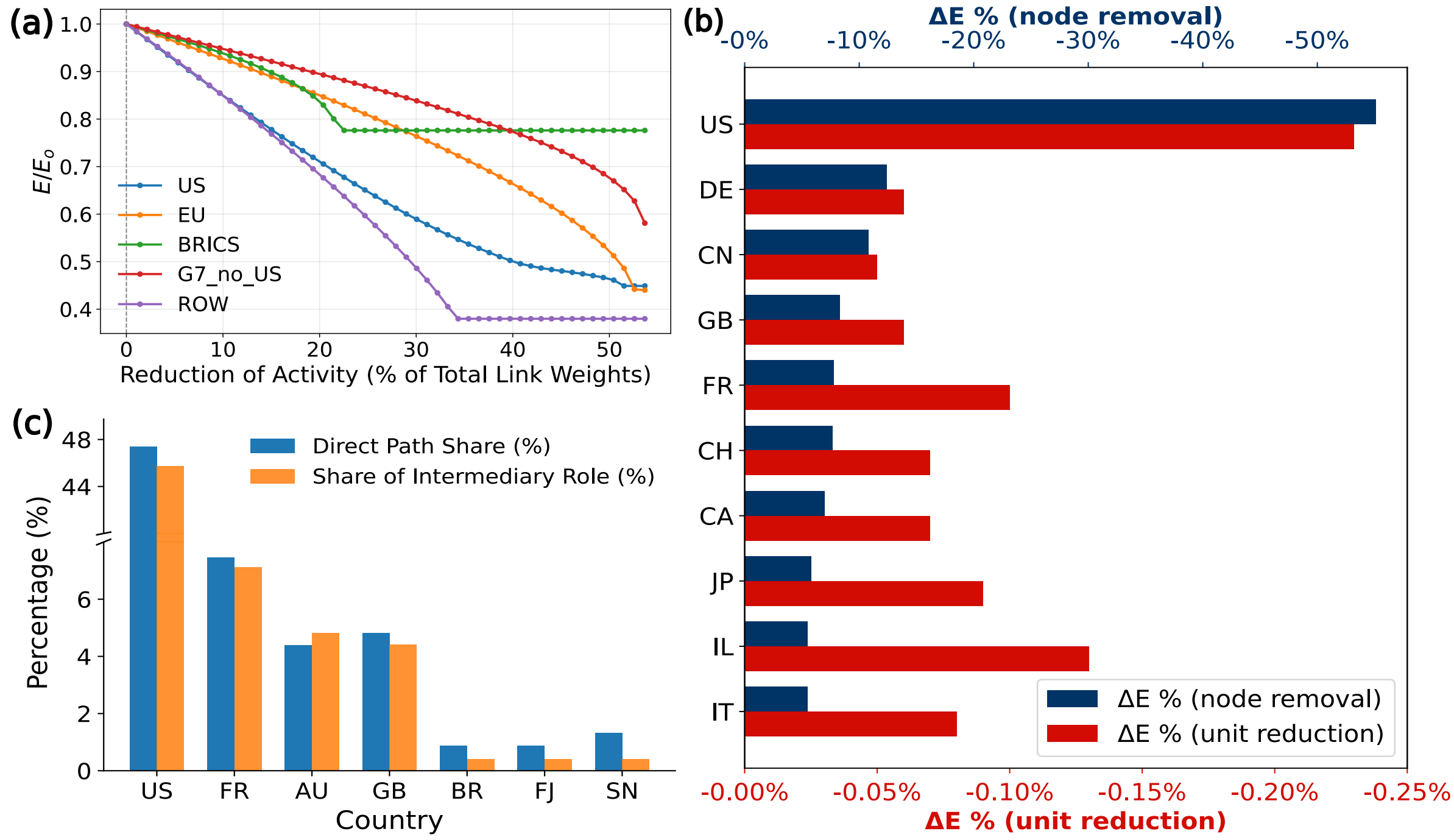

Fig2. Fragility analysis of the cancer science collaboration system based on the aggregated network. (a) Dependence of the global network efficiency on a reduction of activity corresponding to a proportion of total link weights for the US, EU (n=27), BRICS (n=5), G7 excluding the US (n=6), and ROW (n=197). (b) Percentage reduction in global network efficiency after complete removal (blue) of and unit cut (red) to activities of individual countries from the network, shown for the 10 countries with the largest impacts. A unit cut here corresponds to 1% of CN's total activity and is applied through scaling links adjacent to each country. (c) Share of direct (blue bars) and intermediary (orange bars) paths for four leading countries and three smaller countries.

## Compensation Analysis

Having studied the dependency on countries' roles in collaboration, we are next interested in ways how efficiency losses due to activity reductions of individual countries could be compensated for by other actors. For this purpose, we use the conceptional framework outlined in Fig3(a). Having first reduced the activity level of a country, we then freeze its connections and explore upscaling other countries' activities until the original efficiency level is restored. For instance, Fig3(a) proposes scenarios in which the EU, G7 countries (excluding the US) or BRICS countries increase their respective activities to compensate for a loss of US activity.

To quantify the required compensation activity, we hypothesized a 40% reduction in US research activity[2,3] and computed the fraction of activity increase required to re-establish the original efficiency level. Results of the compensation analysis for various country groupings are shown in Fig3(b). We find that the original communication efficiency of the cancer science system could be restored by an additional 123% of EU activity, whereas the required commensurate increases would be 130% for the remaining six G7 countries, 354% for the BRICS, or 144% for the rest of the world combined (ROW). Percentage-wise activity increases seem to suggest that the most reasonable path towards compensation might be through EU or G7 increases of activity.

We also carried out comparisons of the absolute amounts of compensatory activity needed (Supplementary Table 2). Translating activity in the aggregate network from normalised units into the number of grants, papers, patents, or inventions alone per year (Supplementary Figure 7), we find that it would require an additional 100 international grants (totaling $17 million based on median award size), 1451 clinical trials, 26999 publications, or 821 inventions from the EU. For the non-US G7 countries, 98 grants ($5 million), 1422 clinical trials, 26462 publications, or 796 inventions are required. BRICS would require an additional 84 grants ($4 million), 1223 clinical trials, 22748 publications, or 684 inventions. The ROW would need an additional 77 grants ($36 million), 1120 clinical trials, 20857 publications, or 627 inventions.

We find that the overall most efficient compensation schemes, measured in terms of the number of grants, papers, patents, or inventions alone, tend to be the ones where the burden is shared most broadly, i.e., through the ROW. However, if considering median award size of grants, the BRICS and non-US G7 countries are the most cost-efficient options to compensate for the US reduction in funding.

In principle, increases in international collaboration activity could be achieved in two ways. First, countries could increase their total activity by a certain amount. However, as our analysis above shows, the required amounts might be unrealistically large. A second way to improve international activity is by devoting an increased share of existing resources to international collaborative activity. Below, in Fig3(c), for all the separate layers of the value chain and for various country groupings we analyse current and target fractions of resources devoted to international activity required to compensate for a 40% cut in US activity. We note that in the publication, clinical trial, and invention networks, it often requires more than a doubling in the fraction of resources devoted to international activity. For instance, the EU would need an increase from 15% to 37% to restore the invention network, 18% to 34% to repair the publication network, and 17% to 50% for the clinical trial network. In contrast, the increases required to compensate in the grant network are relatively small. For instance, the EU would only need a relatively minor increase from 2.4% to 2.7%. In conclusion, whereas in the case of the publication, clinical trial and invention, the required increases to international collaboration activity seem substantial, repairing collaboration in grant funding only involves relatively minor changes.

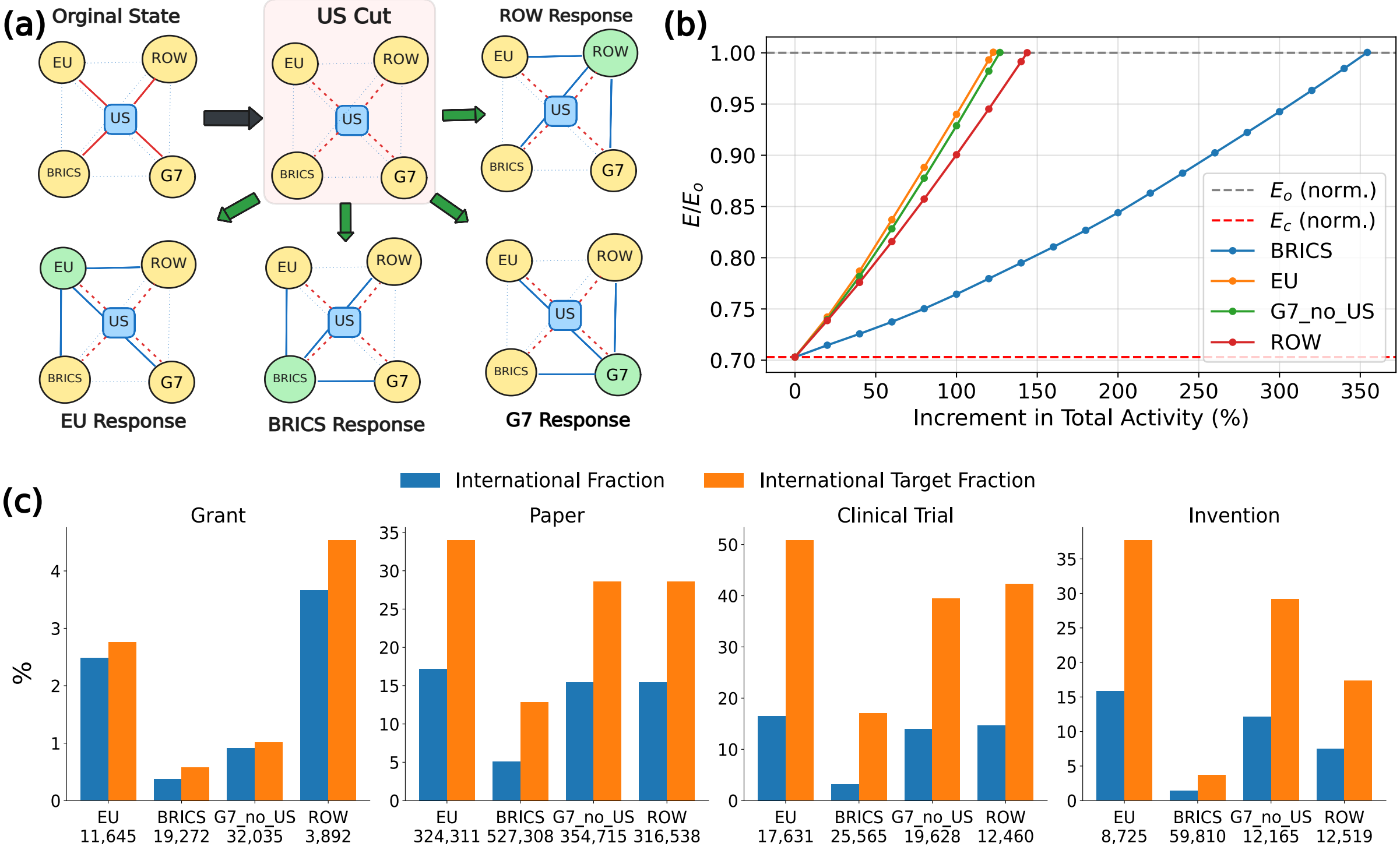

Fig3. Analysis of possible ways how other actors can compensate for a loss of 40% US activity. (a) The compensation framework. In the original network, the US collaborates with the EU, BRICS, G7, and ROW (red solid lines). Then, after the US cut, collaboration between the US and other countries is reduced (red-dashed lines). To restore network efficiency, country groups can respond to the US cut by enhancing their international collaboration intensity (blue solid lines). As indicated, potential response scenarios could be via activity increases of the EU, BRICS, G7, or ROW. (b) Dependency of system-wide relative efficiency on upscaled activity of the EU, BRICS, G7, and ROW after a US cut. Dashed lines give the efficiency levels before (grey) and after (red). (c) Current fraction (blue bars) of international activity and target fraction (orange bars) of international activity required to restore the global efficiency to the original level after a US cut for the grant, publication, clinical trial and invention layers.

## Network Resilience

Having studied how losses due to declines in individual country activity could be compensated by other actors, we next investigate the fragility of the compensated networks to further shocks of the cancer science system. We do this by testing the resilience of the compensated networks to removal of collaborations between pairs of countries. We model this in simulations of random link failures on the aggregated multiplex and evaluating their impact on network efficiency. Fig4(a) visualises results and shows the decrease in network efficiency as a function of the fraction of removed links.

Two main observations stand out. First, we note that the current aggregated collaboration network is substantially more resilient than randomized networks with the same link weight distribution. This is reminiscent of the enhanced robustness of scale-free networks to random attacks and might thus be due to the concentrated US influence.[25] Second, we note that networks with more even distributions of link weights are more resilient.[10,11] We also note that the compensation process spreads originally-US-centred link weights more evenly through the network. Therefore, we find that compensation for a reduction of US activity also tends to lead to an overall gain of system resilience.

To illustrate, we note that the EU compensated network is substantially more resilient than the original network (Fig4(a), green curve). Conversely, the network becomes significantly less resilient in a scenario in which the US compensates for a modelled EU activity cut (Fig4(a), red curve). Similar observations hold for most compensation scenarios that we analysed (Supplementary Table 4). The analysis reveals opportunities in potential compensation mechanisms for US activity, which have the potential to make the overall cancer collaboration ecosystem more resilient.

Following analysis of resilience, we next explore robustness to activity reductions of individual countries. To ensure a fair comparison, we construct compensated networks by picking a fixed amount of a cut and applying this cut on the US, the EU, and BRICS, respectively. Then, we compensate for this fixed cut through scaling up all other unaffected links. Comparing the original and compensated networks, Fig4(b) analyses the differential change in communication efficiency of country groupings to further cuts in their respective activity. We find that compensating for the activity of a country can result either in a loss or gain of resilience. For instance, we note that the compensated US network is more resilient to a further US cut, whereas the compensated EU or compensated BRICS networks become less resilient to further EU or BRICS cuts Fig4(b).

An understanding of these effects can be gained through an analysis of changes in intermediary paths, see Fig4(c, d). In more detail, we count US/EU/BRICS-related intermediary paths gained or lost in the compensated network subject to the cut compared to the original network along with the average path length of the respective paths in Fig4(c). In Fig4(d), we count intermediary paths that are present before and after the cuts and analyse changes in their path lengths.

Results of the path length analysis are shown in Fig4(c). We observe that an EU or BRICS compensation for a US cut has the potential to substitute many formerly US-routed intermediary paths by more efficient intermediaries now routed through the EU or BRICS.

For those intermediaries that still go through the US, around half has a reduced path length, with a larger effect size than the path length increases for the other half (Fig4(d)). In contrast to compensating for a US cut, when compensating for the EU or BRICS, the rerouted paths gained are generally less efficient (Fig4(c)). Notably, a large proportion of paths that still route through the EU or BRICS have seen an increased path length, with a much larger effect size than their decreased-length counterparts (Fig4(d)).

In the current system, the US provides the role of an important intermediary for connections between other countries (Fig Appendix Pie Chart and Fig2(c)). Reducing US activity and upscaling other countries activities re-routes intermediary paths, shortens path lengths, and reduces overall system dependency on the US (see Fig.4(c,d)). Consequently, further US cuts will affect the system to a lesser extent (Fig4(b)), as the modelled compensation activity now has (i) increasingly re-routed paths to other more efficient intermediaries (Fig4(c)) and (ii) improved the efficiency of the unchanged paths Fig4(d). Additionally, reducing EU or BRICS activity tends to reroute paths to more inefficient intermediaries (Fig4(c)), and reduces the efficiency of intermediaries through the EU or BRICS. Therefore, reductions of activity in the EU and BRICS lead to enhanced vulnerability in the compensated systems (Fig4(b)).

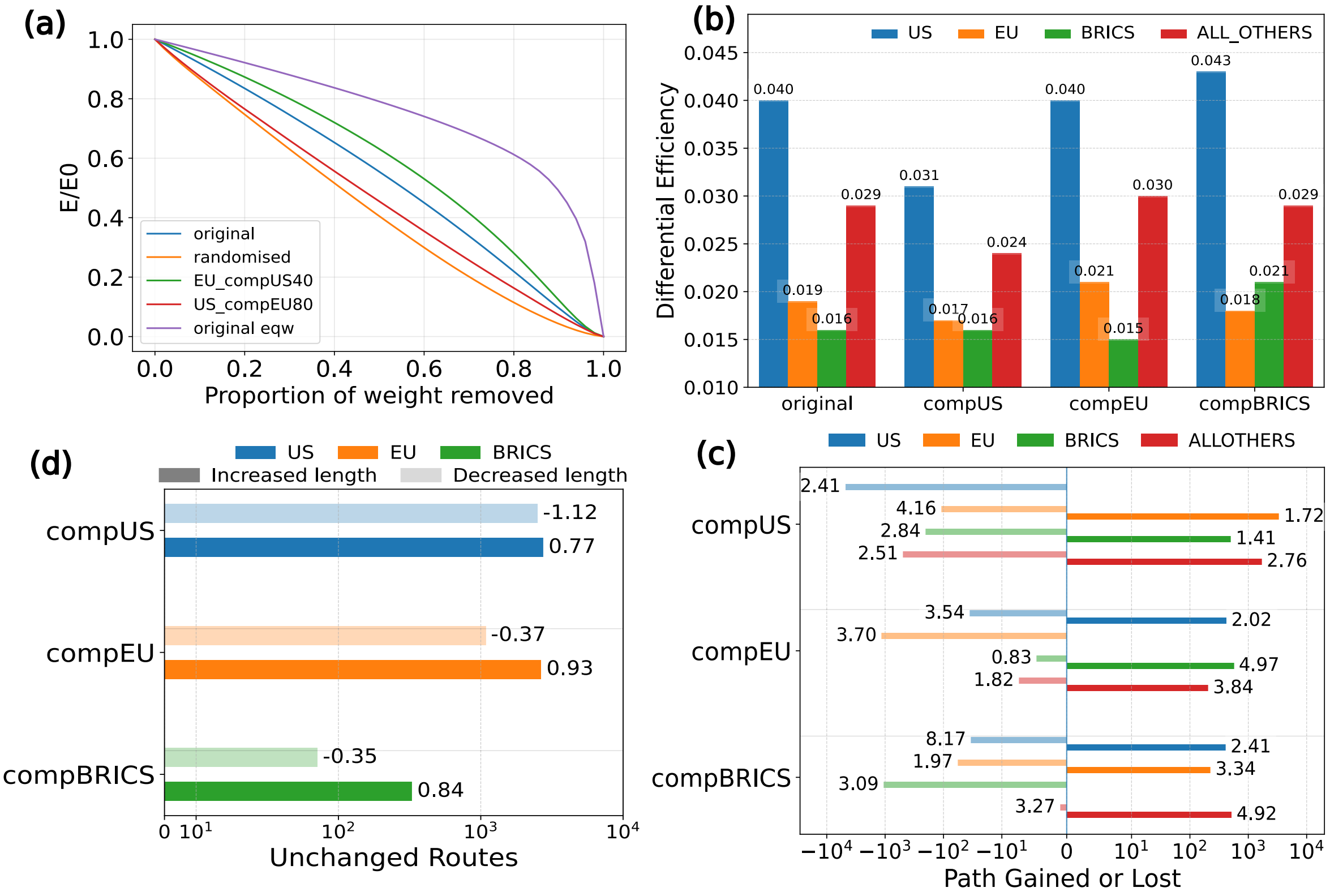

Fig4. Resilience analysis of the 4-layer system. (a) Testing network resilience subject to random link removal for the original aggregated network (blue), and for comparison, an ensemble of networks generated from the aggregated network through link randomization (orange), the aggregated EU compensated network (green), and a network with the same topology as the original network, but with all links having equal link weight (red). Curves show averages and 95% confidence intervals derived from 10,000 simulations of random failure. (b) The reduction in network efficiency of a unit cut corresponding to 1% of CN's total strength on US/EU/BRICS in the original and compensated networks. (c) Counts of US/EU/BRICS-related intermediary paths gained or lost in the compensated network subject to a cut corresponding to 80% of Chinese activity to the US/EU/BRICS compared to the original network. This quantity is picked so that it will not exceed the total link weight of each country or group. Annotated numbers represent the average path lengths. (d) Count of unchanged intermediary paths which have increased (light colours) or decreased (dark colours) path length, with the corresponding average change in path length annotated.

## Discussion

The world systems theory[26,27] divides the world into core and periphery zones with unequal exchange. As our results indicate, the international cancer science ecosystem is no exception – the inequality manifests itself in the ecosystem through not only a core-periphery structure but also through consistent absence of LMICs towards the two 'monetary ends' of the science value chain (i.e. grants and patents). The existing capacity in the research and innovation ecosystem might limit LMICs' ability to acquire sufficient funding from international funders and commercialise the knowledge.[5]

While there is stronger involvement of LMICs in the publication layer, their involvement declines when considering specifically first/last-author co-authorship. This indicates their involvement as being less valued and arguably potential tokenism. These inequalities have profound implications. Researchers in LMICs have less opportunities to generate research income, innovate, and patent their ideas. This increases local institutions' dependence on higher income countries, with potential for 'brain drain' of highly-skilled workers who migrate to richer countries.[28] These may further reinforce the rich core's dominance over LMICs, causing a persistent net welfare loss.

Since 2025, there has been substantial declines in funding for biomedical research in the US[4,29] with disruption to existing and planned research and displacement of many scientists.[30] Our findings indicate that the current international cancer science ecosystem is fragile, and highly affected by disruptions in US research funding, because the US serves as the biggest hub and the most critical provider of bridge for information exchange in the network. In contrast, disruption on other hubs like China has less impact on information exchange.
Our compensation analysis demonstrates that country groups can increase support for international activities to restore the original level of knowledge sharing before the declines in US funding. Percentage wise, compensation seems relatively easier for higher-income countries, but overall, most cost-efficient compensation could be achieved by sharing compensation efforts most broadly.

Our analysis also shows that the proposed compensation strategy can potentially be operationalised without additional resources through increased internationalisation of existing activities. Such a path seems most realistic regarding grant funding, where only relatively minor increases in internationalisation are required.

Many public R&D systems already possess institutional mechanisms for internationalisation, such as the UK's International Science Partnerships Fund, Japan's ASPIRE programme, China's intergovernmental science and technology cooperation programmes, and Germany's DFG international schemes.

China has already been playing an increasingly important role in research and development.[31,32] A recent study has documented a substantial growth of China's early-stage biopharmaceutical development programs between 2015 and 2025.[33] Chinese companies are emerging as leaders in biotech innovation, with partnerships and licensing deals with other countries, which might provide ways towards stronger collaborative links in the future.

We have also demonstrated that compensation for declines in US activity may not only be possible in some scenarios but would also offer distinct benefits in terms of enhanced system resilience. In this regard our analysis shows that paths towards enhanced resilience generally go through a broader sharing of scientific activity in all layers of the value chain. Our research provides a new quantitative analysis for reducing fragility and improving resilience of international cancer science ecosystem through country alliances with existing budget envelopes and maintain the knowledge flows and scientific collaboration.

# Method

## Data Source and retrieval

Grant records related to cancer were retrieved from the Digital Science Dimensions database and selected for inclusion/exclusion based on a mixture of expert judgement and automated classification using machine learning and a large language model (OpenAI's GPT4o-mini), following the procedure outlined in previous work.[5] 107955 awards totaling $51.4 billion from 2016-2023 were considered.

Patent documents with cancer-relevant expert-assigned CPC and ICPR classification code (see appendix for a list of codes) and priority year within 2016-2023 have been retrieved. Patent documents were deduplicated with respect to unique DOCDB simple patent family,[34] defined as one unique invention, by retaining only the earliest-filed patent documents under each family. As a result, 113937 unique patent families were considered.

Clinical trial records have been retrieved through title and abstract search of previously curated keywords by domain experts with a start year of the trial in 2016-2023. Each record carries a 'conditions' field that contains the focal disease and categorical tags including Fields of Research (FOR), Research, Condition, and Disease Categorization (RCDC), Health Research Classification System Health Category (HRCS_HC), Common Scientific Outline (CSO), Cancer Type (CT), etc. We included trails with a 'conditions' field having a cancer related keyword and the trials with a non-empty CSO or CT field, or '3211 Oncology and Carcinogenesis' in the FOR field, or 'Cancer and neoplasms' in the HRCS_HC field, or cancer specific topics in the RCDC field. As a result, 89810 clinical trial records were included for analysis in this study.

Papers have been retrieved through title and abstract search of previously curated keywords by domain experts (see appendix "Keywords for data retrieval"). These were first filtered in terms of document type, where we only included citable documents, i.e., Conference Paper, Research Article, Research Chapter, and Review Article, as per Dimensions definition.[35] As with clinical trial records, we included those papers with a non-empty CSO or CT field, or '3211 Oncology and Carcinogenesis' in the FOR field, or 'Cancer and neoplasms' in the HRCS_HC field, or cancer specific topics in the RCDC field. As a result, 1779449 papers were included for analysis in this study.

## Fractional Counting and Network Construction

As our primary interest is on country level collaborations, we construct networks with nodes representing countries and weighted edges corresponding to collaboration intensity between these countries. We adopt fractional counting[24] so that each country pair is attributed with an equal shares of one grant, paper, clinical trial, or patent. Compared to a full counting approach,[36] the fractional counting approach prevents overcounting due to very large collaborations and ensures a fair cross-country comparison.[24] Note that records with a single affiliated country are counted as self-links, which were excluded from the network construction.

Following this procedure, weighted collaboration networks have been constructed using information on grant co-investigation, paper co-authorship, clinical trial co-investigation, co-invention, and patent co-ownership, resulting in a 5-layer multiplex covering 233 countries and territories (see countries and territories in the appendix). For a summary representation of overall collaboration intensity for inequality analysis, we first normalise individual layers of the 5-layer multiplex to have identical average link weight and then aggregate. Slightly differently, to prevent double counting of patent layers (represented through invention and ownership) in the fragility, compensation, and resilience analysis, we also considered an aggregated 4-layer multiplex based only on grant, paper, clinical trial and invention layers.

The total activity of a country is defined as the sum of domestic and international activity, where domestic activity refers to the weights of self-links that were omitted from network construction and international activity refers to the node strength.

## Network Decomposition

We study the core membership by network decomposition, i.e., we define link-weight cutoffs and analyse the largest connected components of the networks composed of only links with weights that exceed the cutoff threshold.

## Measure of Network Function and Path Counting

We assess system function based on network efficiency,[10,11,22] which has been widely adopted to track the functional performance in physical, biological, transportation, and social networks.

We define the shortest path length as $d_{ij} = \min_{h_1,h_2,..,h_k} (1/w_{ih_1} + 1/w_{h_1h_2} + \cdots + 1/w_{h_{k-1}h_k} + 1/w_{h_kj})$ with $h_1, h_2, .., h_k$ being the intermediary nodes on the shortest path between node i and j, and $w_{mn}$ the link weight of the link (m,n). The network efficiency is then defined as $E = \sum_{i\neq j} \frac{1}{d_{ij}} / n(n-1)$, where n is the number of nodes and $d_{ij}$ is the length of the shortest path between nodes i and j.

Note that, due to the granularity of link weights, in practice each shortest path is unique in the networks studied. When counting shortest paths that route through multiple countries, we attribute equal fractions of the path to each country it touches.

## Compensation Analysis

We study possible responses from various groups of countries (EU, BRICS, G7_no_US, rest of the world) that could restore network efficiency after a reduction of the US activity in the network. To study shocks of different sizes, we implemented a 40% and an 80% reduction.[2–4] The activity reduction is operationalized through downscaling the weights of all US related links. Compensation is then implemented through upscaling non-US link weights of compensating countries. We keep upscaling the activity of compensating countries until the original network efficiency is restored. We carried out compensation analysis for the aggregated network and individual network layers (Supplementary Table 2). For comparison, compensatory efforts from each country or group towards cuts in EU and BRICS were reported in Supplementary Table 2 which compares analysis of percentage-wise and absolute amount reductions.

### Resilience Analysis

To understand whether compensated networks exhibit more resilience to shocks, we test the network resilience through tracking the efficiency decay in simulations of random link removal.[10,11,22] For each network, we compute the dependence of the average efficiency on the percentage of total link weight removed from 10,000 simulations and quantify overall fragility by the area under the curve. For comparison, we perform these simulations also on two types of reference networks. The first is an ensemble of networks generated from the aggregated network through link reshuffling.[10,11] The second is a network with the same topology as the original network, but with all links having equal link weight.

In the compensated network shortest paths may be routed. Three scenarios can occur. First, paths through a country can be rerouted through other countries, resulting in a loss of intermediary paths for the reference country. Second, paths previously routed through another country can now be rerouted through the reference country, contributing to a gain of intermediary path. Lastly, paths can remain unaltered.

We count the number of paths gained or lost in the first two scenarios and report the average path lengths for these paths. For unaltered paths, we keep track of the average change in path length, reporting separate statistics for paths that become longer or shorter.

### Data Availability

The original data that support the findings of this study are available from Digital Science Dimensions but restrictions apply to the availability of these data, which were used under license for the current study, and so are not publicly available. Datasets and codes generated during this study are however available from the authors upon reasonable request and with permission of Digital Science Dimensions.

## Acknowledgement

The authors acknowledge the use of the IRIDIS High Performance Computing Facility, and associated support services at the University of Southampton, in the completion of this work.

## Author Contribution

M.B. and A.D. conceived the study. M.G.H, A.D., and B.Z. collected the data. A.D. carried out data analysis and wrote the manuscript. R.A., M.G.H, B.Z., and M.B. critically revised the manuscript. M.B. and M.G.H supervised the project.

## Competing Interest

The authors declare no competing interests.

## Funding

This work did not receive any funding.